\documentstyle{mn}
\input epsf

\begin{document}
\journal{Preprint astro-ph/9907224} 
\title[Hydrodynamical Sunyaev--Zel'dovich effect simulations] 
{Hydrodynamical simulations of the Sunyaev--Zel'dovich effect} 
\author[A.~C.~da Silva et al.]{Antonio C.~da Silva,$^1$\thanks{Address from 1st 
January 2000: Astronomy Centre, University of Sussex, Brighton BN1 9QJ} 
Domingos Barbosa,$^1$ Andrew R.~Liddle$^{1\mbox{\LARGE $\star$}}$\cr and
Peter A.~Thomas$^2$\\ 
$^1$Astrophysics Group, The Blackett Laboratory,
Imperial College, London SW7 2BW\\ 
$^2$Astronomy Centre, University of Sussex, Brighton BN1 9QJ}

\maketitle

\begin{abstract}
With detections of the Sunyaev--Zel'dovich (SZ) effect induced by
galaxy clusters becoming routine, it is crucial to establish accurate
theoretical predictions.  We use a hydrodynamical $N$-body code to
generate simulated maps, of size one square degree, of the thermal SZ
effect. This is done for three different cosmologies --- the
currently-favoured low-density model with a cosmological constant, a
critical-density model and a low-density open model. We stack
simulation boxes corresponding to different redshifts in order to
include contributions to the Compton $y$-parameter out to the highest
necessary redshifts. Our main results are:
\begin{enumerate}
\item ~The mean $y$-distortion is around $4 \times 10^{-6}$ for low-density 
cosmologies, and $1 \times 10^{-6}$ for critical density. These are below 
current limits, but not by a wide margin in the former case.
\item ~In low-density cosmologies, the mean $y$-distortion is
contributed across a broad range of redshifts, with the bulk coming
from $z \la 2$ and a tail out to $z \sim 5$. For critical-density
models, most of the contribution comes from $z < 1$.
\item ~The number of SZ sources above a given $y$ depends strongly on
instrument resolution. For a one arcminute beam, there is around 0.1 
sources per square degree with $y > 10^{-5}$ in a critical-density
Universe, and around 8 such sources per square degree in low-density
models. Low-density models with and without a cosmological constant
give very similar results.
\item ~We estimate that the {\sc Planck} satellite will be able to see
of order 25000 SZ sources if the Universe has a low density, or
around 10000 if it has critical density.
\end{enumerate}
\end{abstract}


%%%%%%%%%%%%%%%%%%%%%%%%%%%%%%%%%%%%%%%%%%%%%%%%%%%%%%%%%%%%%%%%%%%%%%

\section{Introduction}

The Sunyaev--Zel'dovich (SZ) effect (Sunyaev \& Zel'dovich 1972, 1980;
for reviews see Rephaeli 1995 and Birkinshaw 1999) is the change in
energy experienced by cosmic microwave background photons when they
scatter from hot gas, especially that in galaxy clusters.  It comes in
two forms.  The dominant contribution is the thermal SZ effect, 
the gain in energy acquired from the thermal motion of the gas
which is commonly at a temperature of tens of millions of degrees in
clusters. It is described by the Compton $y$-parameter, also known as the 
$y$-distortion. The kinetic SZ effect is the Doppler shift arising from the
bulk motion of the gas.

Since the first claimed detection by Parijskij (1972), and the
more recent pioneering work of Birkinshaw, Hughes \& Arnaud (1991) and
Birkinshaw \& Hughes (1994), detections of the thermal SZ effect from
clusters have become routine.  A number of instruments have become
capable of making two dimensional maps, including the Ryle telescope,
centimetre receivers on BIMA and OVRO, Viper, SuZie, SEST, PRONAOS and
Diabolo.  A blank-field survey has been
proposed (Carlstrom et al.~1999), and eventually the {\sc Planck}
satellite will produce an all-sky catalogue likely to contain
thousands of SZ sources.

Because clusters of galaxies are in the tail of the Universe's mass
distribution function, their number density is highly sensitive to the
growth rate of density perturbations, which is governed primarily by
the density parameter $\Omega_0$, but also influenced to a lesser
extent by the density, $\Omega_\Lambda \equiv \Lambda/3H^2$,
contributed by a cosmological constant $\Lambda$. Korolev, Sunyaev \&
Yakubtsev (1986) were the first to point out the interest of the SZ
cluster source counts as a cosmological probe.  Since then, a plethora
of authors using a variety of techniques have discussed how the
cosmological cluster evolution could be revealed by the SZ effect
(Cole \& Kaiser 1988; Cavalieri, Menci \& Setti 1991; Markevitch et
al.~1991, 1992, 1994; Bartlett \& Silk 1994; Barbosa et al.~1996; Eke,
Cole \& Frenk 1996; Aghanim et al.~1997) and how gas evolution could
mask it (Bartlett \& Silk 1994; Colafrancesco et al.~1994, 1997).

Where the above-mentioned authors used extrapolations based on the
known cluster X-ray luminosity functions to estimate the source
counts, the preferred tool was without doubt the Press--Schechter
(1974) approximation.  Although in excellent agreement when compared
to previous numerical simulations, its ansatz usually assumes that
clusters can be modelled by some spherical, and usually isothermal,
matter distribution.  The maturity of this field now demands a
sophisticated approach to the theoretical modelling of the effect,
using hydrodynamical $N$-body simulations.  The pioneering study of
this kind, using a crude `sticky particle' method, was by Thomas \&
Carlberg (1989) who made maps with one arcminute resolution.  Cen \&
Ostriker (1992) used an Eulerian hydrodynamic code in order to study
the mean $y$-distortion, and this was followed by Scaramella, Cen \&
Ostriker (1993) who used these simulations to make maps,
using the standard Cold Dark Matter model which is however now
excluded by observations.  Persi et al.~(1995) used hydrodynamic
simulations as part of a semi-analytic calculation of the anisotropy
power spectrum from the SZ effect.  Hydrodynamic simulations have also
been used to examine the properties of individual clusters (Metzler
1998), in particular seeking the relationship between $y$ and the
cluster mass $M$.  In this paper, we make simulated maps of the SZ
effect, using state-of-the-art SPH hydrodynamical simulation
techniques. This is done for three currently-popular cosmologies,
enabling a comparison between them. These maps are of particular
importance to simulate the SZ sky that near-future experiments will
observe.

\section{The simulations}

Viable cosmological models must be able to reproduce the number
density of clusters seen at the present epoch, which is quite well
constrained observationally. However, in a map of a given angular size
one expects most SZ clusters to be at quite significant redshifts,
where observations have yet to constrain the number of clusters;
consequently predictions become quite model dependent and one needs to
consider several different models. Primarily, the number of clusters
depends on the density parameter $\Omega_0$, with less prominent
dependence on the cosmological constant $\Lambda$; assuming gaussian
initial perturbations these are the only significant quantities as
they determine the growth rate of density perturbations and hence the
epoch of cluster formation.

We therefore consider three models, all from the cold dark matter
(CDM) family. In each case the power spectrum was that of CDM with
shape parameter $\Gamma=0.21$, and the normalization $\sigma_8$ was
chosen to ensure good agreement with the present cluster number
density (see e.g.~Viana \& Liddle 1999). The baryon density
$\Omega_{{\rm B}}$ was chosen to agree with nucleosynthesis for a
reasonable choice of the Hubble parameter $h$; we used 
$\Omega _{{\rm B}}h^{2}=0.02$, with $h=0.71$ for the low-density cosmologies and 
$h=0.50$ for the critical-density model. As cooling is not included, the 
simulations scale and for them $h$ need not be specified. However, since the
$y$-distortion parameter is an integral along the line of sight, $h$ is 
required. How the derived $y$ varies with $h$ depends on how one chooses to 
scale the baryon density. If $\Omega_{{\rm B}}$ were fixed (preserving the 
gravitational forces), then the electron density $n_{{\rm e}}$ scales as $h^2$ 
and the scaling is $y \propto h$. However if we regard the gravitational effects 
of baryons as negligible, we could instead keep $\Omega_{{\rm B}}$ at the 
nucleosynthesis value and then $n_{{\rm e}}$ is independent of $h$ and the 
scaling is $y \propto h^{-1}$. 

The models we use are
\begin{itemize} 
\item $\Lambda$CDM: a low-density model with a flat spatial geometry and 
$\Omega_0 = 0.35$ and $\Omega_\Lambda = 0.65$.
\item $\tau$CDM: a critical-density model with $\Omega_0 = 1$ and 
\mbox{$\Omega_\Lambda = 0$}.
\item OCDM: a low-density open model with $\Omega_0 = 0.35$ and
$\Omega_\Lambda = 0$.
\end{itemize}
Current observational prejudice is in favour of the low-density flat
model, and where we display results only for a single cosmology that
is the one chosen.

The simulations were carried out using the public domain Hydra code
(Adaptive P$^3$M-SPH: Couchman, Thomas \& Pearce 1995).  In each case
the box-size was $150\,h^{-1}$ Mpc, and
equal numbers of dark matter and gas particles were used. The number of 
particles $N$ was
chosen so as to keep the mass of the dark matter particles, $m_{{\rm
dark}} = 4.45 \times 10^{11} \, h^{-1} M_{\odot}$, the same in each
simulation. The same realization of the power spectrum was used in
each case, though $\tau$CDM, having more particles, better samples the
small-scale power.  The softening was set at 40$\,h^{-1}$kpc.  The other
simulation parameters are given in Table~\ref{tab:simparams}. The effective 
resolution for the baryonic component is 32 times the gas particle mass, about 
$10^{12} M_\odot$.

The closest antecedent to this work is the excellent paper of
Scaramella et al.~(1993). They carried out a large number of separate
simulations, but because of the poorer resolution computationally
accessible at that time, these had sizes ranging from just $4 \,
h^{-1}$ Mpc to $64\,h^{-1}$ Mpc. Because of this, their simulations
were significantly lacking in large-scale power and there were too few
rich clusters. Although we only carry out a single simulation for each
cosmology, this is of much greater volume and with higher
resolution. In total, the volume of the Universe we simulate is
comparable to theirs, but our large boxes contain all the necessary
large-scale power, and allow a number of statistically independent
lines of sight to be traced through them. Also, they only simulated the
Standard CDM cosmology, which has since been shown to be a poor fit to
observations; our three models provide good fits to current
large-scale structure observations, especially the nearby cluster
number density.

\begin{table}
\caption{The cosmological and numerical parameters for the simulations.}
\label{tab:simparams}
\begin{tabular}{ccccccc}
Model & $\Omega_0$& $\Omega_{\Lambda,0}$ & $\Omega_{{\rm B},0}$&
$\sigma_{8}$ & $N/2$ & $m_{{\rm gas}}/h^{-1} M_{\odot}$\\ 
\hline
$\Lambda$CDM& 0.35& 0.65& 0.038& 0.92& $90^3$& $4.85\times10^{10}$\\
$\tau$CDM& 1& 0& 0.08& 0.56& $128^3$& $3.55\times10^{10}$\\ OCDM&
0.35& 0& 0.038& 0.80& $90^3$& $4.85\times10^{10}$
\end{tabular}
\end{table}

\section{The map-making technique}

\subsection{SZ basics}

The thermal SZ effect in a given direction is computed as a line
integral, which gives the Compton $y$-parameter
\begin{equation}
\label{eq1}
y = \int {{k_{{\rm B}} \sigma_{{\rm T}}} \over {m_{{\rm e}} c^2}}\, \,
	T_{{\rm e}} \, n_{{\rm e}}\, dl \,.
\end{equation}
In this expression $T_{{\rm e}}$ and $n_{{\rm e}}$ are the temperature
and density of the electrons, $\sigma_{{\rm T}}=6.65\times 10^{-25}\,
{\rm cm}^2$ the Thomson cross-section, $c$ the speed of light and
$m_{{\rm e}}$ the electron rest mass. Usually, observers prefer to
quote either the ``SZ flux'' at frequency $\nu$ on a line of sight,
\begin{equation}
S_{\nu}=j(x)\, y \,,
\end{equation}
where $x=h\nu/kT_{\gamma}$ is the dimensionless frequency and the
function $j(x)$ is the well-known frequency dependence of the thermal
SZ effect accounting for the decrement or increment in flux with
respect to the mean CMB background flux \cite{sunzel72,birkinshaw99},
or to quote the temperature fluctuation given by
\begin{equation}
\frac{\Delta T}{T}=y \, \left[\frac{x}{\tanh (x/2)} -4 \right] \,.
\end{equation}
In the long-wavelength limit $x \ll 1$ (the Rayleigh--Jeans portion of
the spectrum), this reduces to $\Delta T/T \simeq -2y$.

The simulation output is a set of particles with positions, velocities
and temperatures. Each particle occupies a volume with a radius
proportional to its SPH smoothing length, $h_{\rm i}$. Inside this volume we
choose a mass profile given by $m_{{\rm gas}}W({\bf r}-{\bf r}_{\rm
i}, h_{\rm i})$, where ${\bf r}_{\rm i}$ is the position of the particle's
centre, $m_{{\rm gas}}$ is the mass of the gas particle and 
$W$ is the normalized spherically--symmetric smoothing kernel adopted
in the simulations
\begin{equation}
\label{kernel}
 W(x,h_{\rm i})={1 \over {4\pi h_{\rm i}^3}}\left\{ \matrix{4-6x^2+3x^3\;,\quad 
0\le x\le 1\hfill\cr 
 {(2-x)}^3\;,\quad \quad \quad \kern 1pt \,\,1<x\le 2\hfill\cr
  0\;,\quad \quad \quad \kern 1pt \quad \quad \;\;\;\;\,x>2\hfill\cr } \right. 
\end{equation}
where $x=\left| {{\bf r}-{\bf r}_{\rm i}} \right| /h_{\rm i}$. In order to 
evaluate
$y$ in a single map pixel, we discretize the SZ integral into cubes of volume 
$V$ along the line of sight. If $A$ is the pixel area, then equation~(\ref{eq1}) 
can be 
rewritten as 
\begin{eqnarray}
\label{eq2}
y &=&{{k_{{\rm B}} \sigma_{{\rm T}}} \over {m_{{\rm e}} c^2}} \, 
	      \frac{1}{A} \, \int \, T_{{\rm e}} n_{{\rm e}} \, dV
	      \nonumber \\
          &=&{k_{{\rm B}} \sigma_{{\rm T}} \over {m_{{\rm e}} c^2}} \, 
              \frac{V}{A} \, {0.88 \, m_{{\rm gas}} \over {m_{{\rm p}}}}
              \sum\limits_{\alpha} \sum\limits_{\rm i} T_{\rm i}W(\left| 
              {{\bf r_\alpha}-{\bf r}_{\rm
	      i}} \right| ,h_{\rm i}) \,,
\end{eqnarray}
where the $i$ sum runs over all particles which contribute to the pixel
column, and the $\alpha$ sum is over the line of sight cubes. Here, $T_{\rm i}$ 
is the 
temperature of the gas particles,
$m_{{\rm p}}$ is the proton mass, and the factor 0.88
gives the number of electrons per baryon, assuming a 24 per cent
helium fraction and complete ionization in the regions of interest
(guaranteed as regions have to be hot to contribute significantly to
the SZ signal). The ratio of ionized electrons per baryon will vary
modestly as metallicity builds up, but the effect is negligible in
this context.

\subsection{Map making}

In order to make a realistic map, one must go to high enough redshift
to ensure that all contributions are included. The brightest SZ
sources tend to be quite nearby, but our aim is to make maps down to a
low threshold, $y \ga 10^{-6}$, and with high angular resolution. The
contribution from high redshifts dies off because structure formation
is less advanced, but this is partly offset by two related
effects. For a given mass, high-redshift objects are both hotter and
more concentrated, as the Universe was smaller when they formed. For
instance, the spherical collapse model predicts that for a given mass
the temperature goes as $T \propto (1+z_{{\rm vir}})$, while the
physical radius is smaller by a factor $1/(1+z_{{\rm vir}})$, where
$z_{{\rm vir}}$ is the redshift of virialization. In combination,
these effects mean that quite low mass objects can be seen at high
redshift provided the angular resolution is high enough.

\begin{figure}
\centering \leavevmode\epsfysize=16cm \epsfbox{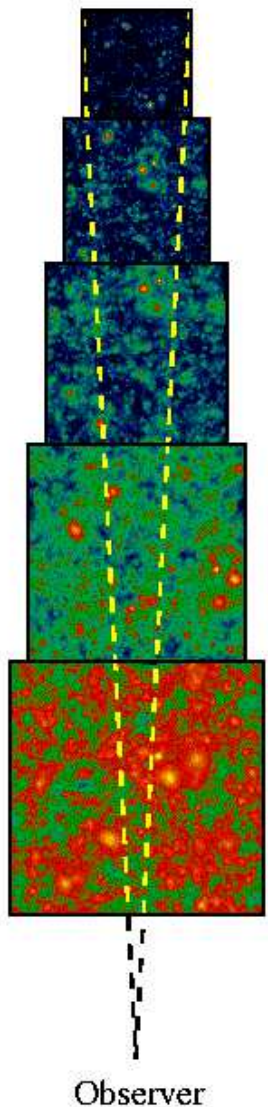}\\
\caption[Figure1]{\label{f:stack} A schematic of the stacking of
simulation boxes to make a map, in physical coordinates. Our actual
stackings are of around 40 boxes. The observer is located at the
centre of the redshift zero box, which we don't include. Notice that
only a small fraction of the volume of the nearby boxes contributes to
the maps.}
\end{figure}

To go to the necessary redshift, we stack simulation boxes in a line
extending to high redshift, as done by Thomas \& Carlberg (1989) and
Scaramella et al.~(1993). Computational resources do not allow us to
make separate simulations for each of the required boxes, but we are
able to use outputs of a single simulation across a wide range of
redshifts. This strategy is clearly not ideal, because the simulation
boxes are not independent realizations, though we aim to minimize the
relation between boxes by randomly translating (using the periodic boundary
conditions of the simulation), reflecting and rotating the boxes
before stacking. Further, the SZ sources we are seeking are
primarily a short-scale phenomenon and the compromise is much less
severe than were one aiming to compute, for example, a correlation
function. Also, the map dimensions are determined so that the most
distant considered box subtends close to the entire map area, which means that
by the time one gets to the nearby boxes only a small fraction of
their volume is being sampled so the nearest boxes are in effect
independent --- see Figure~\ref{f:stack}.

The redshifts of the simulation outputs were arranged so that the
boxes fit together; our stackings include 48, 33 and 39 boxes for the
flat, critical-density and open cosmologies respectively.  We select a
map size which is exactly one degree on a side, and include all boxes
close enough to subtend an angular scale greater than this.  We begin
at redshift greater than 15; however as we will see the contribution from
the boxes is negligible until redshifts around 5 for low-density
models and even less with critical density. For each box, the angular
diameter distance $d_A$ is computed, to indicate which fraction of the
box will contribute to the map; for the most distant ones almost the
entire volume contributes, while for the nearest almost none does.
Within each box we use a small-angle approximation, computing an SZ
map for that individual box by projecting onto a plane located at the
centre of the box.  These maps are made with $300 \times 300$ pixels,
corresponding to an angular resolution of 0.2 arcminutes.  The final
map is obtained by summing together the individual maps.  The
contribution from distant boxes is small because the material within
them is not hot enough to give a significant effect, while that from
nearby boxes is small because such a small volume within them is being
probed that there is little chance of encountering a cluster.

Although we only possess one simulation for each of the three
cosmologies, we are able to make several maps from each by choosing a
different sequence of random translations, rotations and reflections
for each box as it is stacked. We make 30 maps for each cosmology.
While these maps are not truly independent, they give some flavour of
the scatter were one to look to different regions of the sky. For
example, in some realizations, by chance an SZ-bright nearby cluster
appears while in most there are no prominent nearby clusters. Note
that for the box at redshift one, which is around the mean redshift of
contribution for the low-density flat cosmology, only 6 per cent of
its volume contributes to the map, so separate map realizations are
more independent than one might naively expect. Indeed, in the
$\Lambda$CDM cosmology one has to go to $z = 1.67$ for the accumulated volume
contributing to the maps to reach the simulation volume.

In order to give results roughly corresponding to different instrument
configurations, having made the maps we then smooth them with
gaussians of various widths.

\section{Results}

\begin{figure}
\centering \leavevmode\epsfysize=7cm \epsfbox{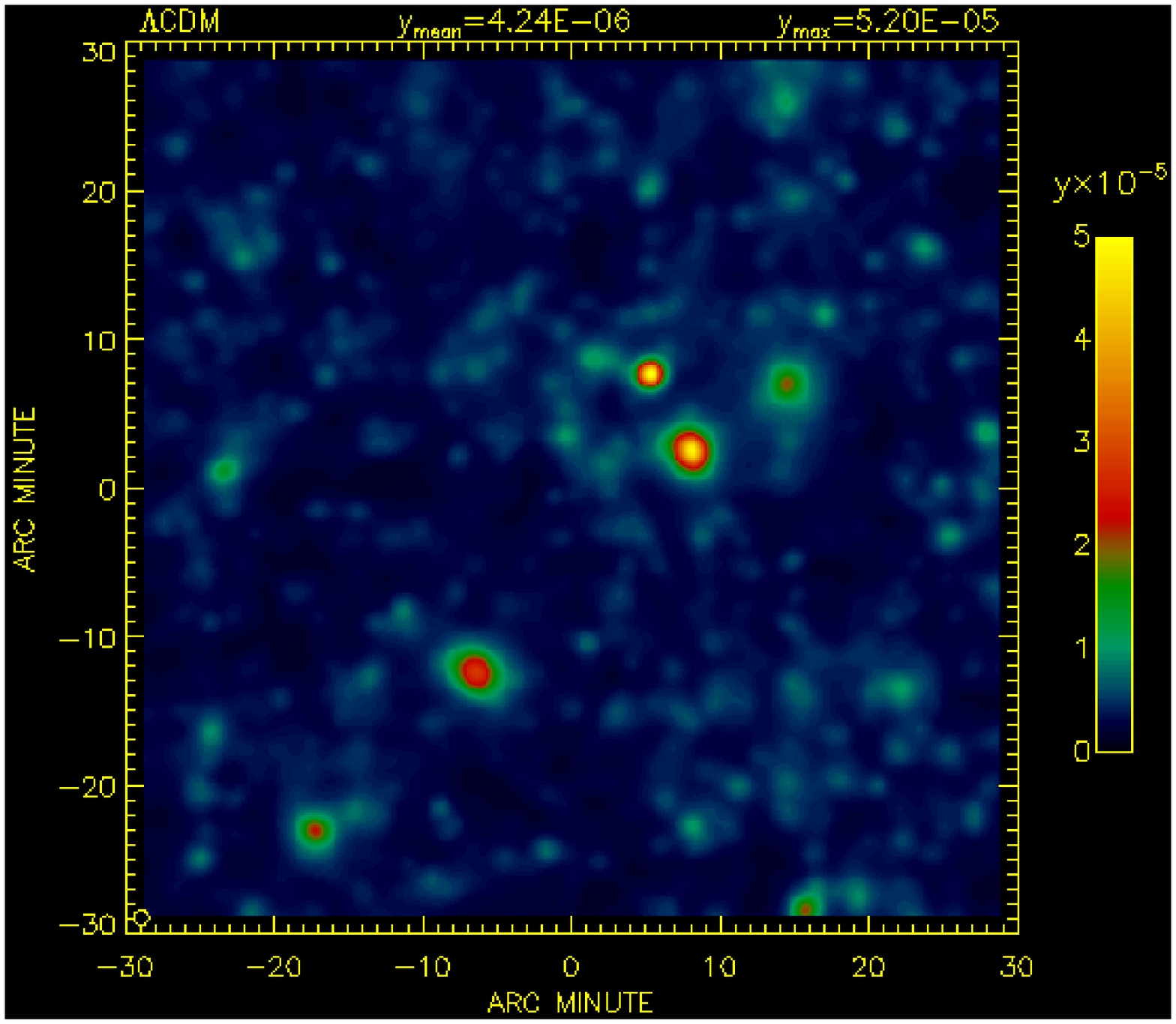}\\ %FLAT
\vspace*{0.15cm}
\centering \leavevmode\epsfysize=7cm \epsfbox{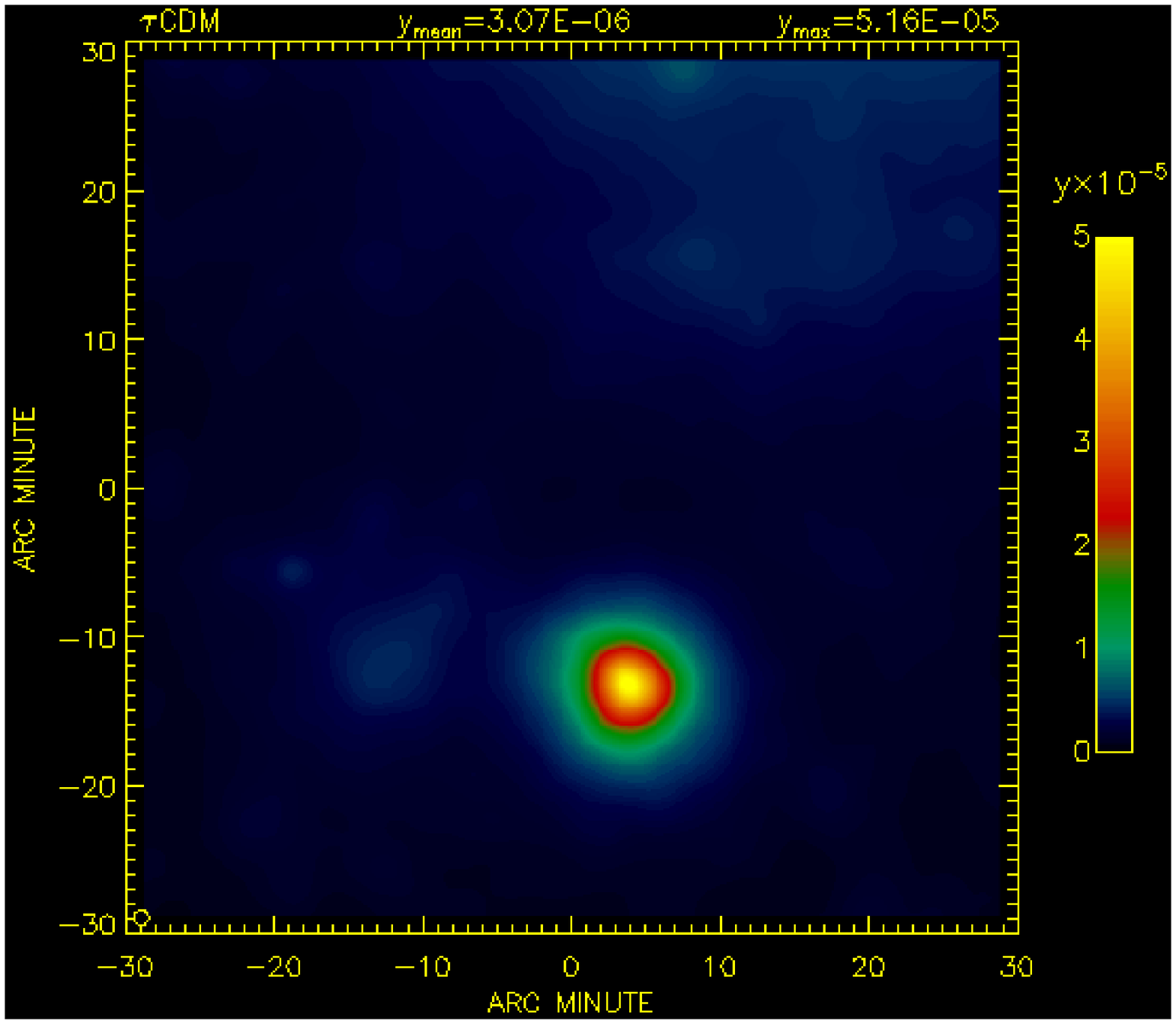}\\ %CRIT
\vspace*{0.15cm}
\centering \leavevmode\epsfysize=7cm \epsfbox{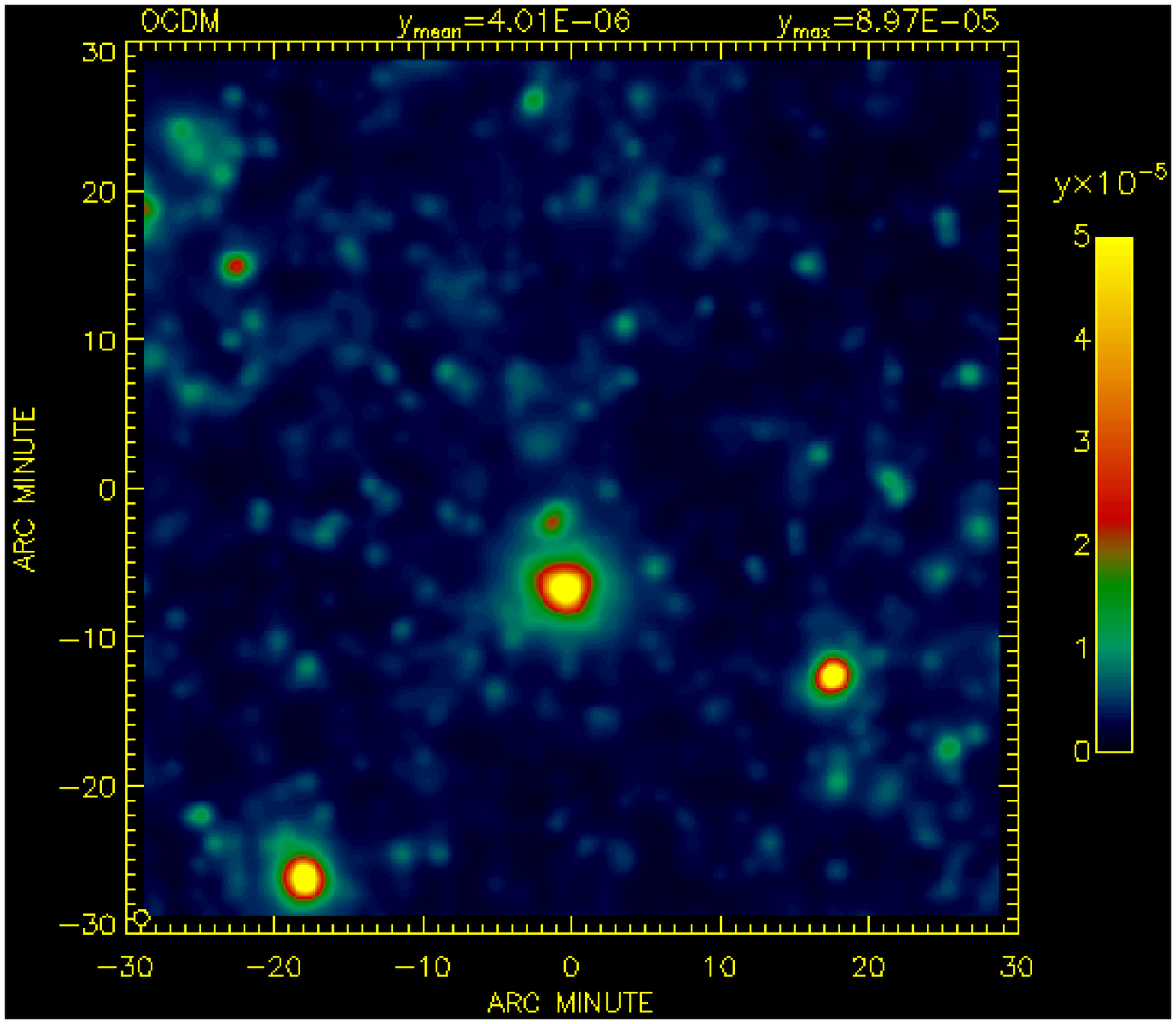}\\ %OPEN
\caption[Figure2]{\label{f:maps} Simulated thermal SZ maps, showing
the $y$-parameter in a patch of one square degree. In each case, the
original map was convolved with a gaussian beam profile of FWHM =
1\arcmin. From top to bottom they are $\Lambda$CDM, $\tau$CDM and
OCDM. The small circles in the bottom left of each image indicate the
smoothing scale.}
\end{figure}

Typical example maps in each cosmology are shown in
Figure~\ref{f:maps}, with the same colour scale in each.\footnote{A
more extensive selection of colour maps, along with animations showing
the contribution from each redshift, can be found at {\tt
http://star-www.cpes.susx.ac.uk/$\sim$andrewl/sz/sz.html}} 
The original maps have
been smoothed using a gaussian with a full-width half-maximum (FWHM)
of 1\arcmin, comparable to the angular resolution of the best existing
experiments.  The visual appearance of the two low-density maps is
rather similar, with several obvious bright spots, corresponding to
clusters, as well as large numbers of fainter sources which
collectively add up to give the mean distortions quoted below. Plotted
with the same colour scale, the critical-density map clearly shows
significantly less structure, which is due to the absence of
high-redshift structures as compared to the low-density cases. Note that this 
particular critical-density map features a nearby large cluster; only about ten 
per cent of realizations for critical density show such a feature.

\subsection{The mean distortion}

The cumulative effect of hot gas, averaged over directions, gives rise to a
mean $y$-distortion across the sky. We see from the maps that,
especially in the low-density cases, there is a significant distortion
along a large fraction of lines of sight. The best observational limit on the 
mean
$y$-distortion comes from the COBE--FIRAS experiment, which sets a 95
per cent upper limit of $y_{{\rm mean}} < 1.5 \times 10^{-5}$
\cite{fix96} for the distortion averaged over a large region of the
sky. All of our cosmologies are below that; averaged over the 30
separate maps made for each cosmology we find
\begin{itemize}
\item $\Lambda$CDM: $y_{{\rm mean}} = 3.9 \times 10^{-6}$
\item $\tau$CDM: $y_{{\rm mean}} = 1.3 \times 10^{-6}$
\item OCDM: $y_{{\rm mean}} = 3.3 \times 10^{-6}$
\end{itemize}
where the statistical uncertainty is much less than systematic
uncertainties from the method.  However the values for the low-density
cosmologies are not far below the current limit.\footnote{There is
also a limit on {\it rms} fluctuations in the $y$-parameter on the
7\degr\ scale, from combining FIRAS with the COBE--DMR experiment
(Fixsen et al.~1997). This is $\Delta y < 3 \times
10^{-6}$ (95 per cent confidence); our maps are too small to allow a
direct comparison, but none of our models are likely to violate this.}
Note also that our simulations do not include non-gravitational
heating which may raise the mean value; this is discussed further
later.

\begin{figure}
\centering \leavevmode\epsfysize=11.4cm \epsfbox{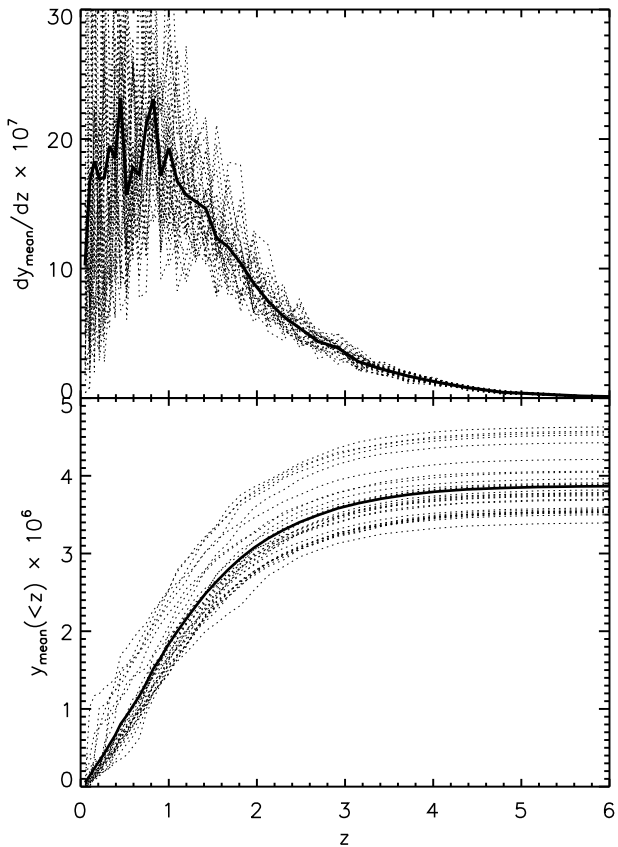}\\
\caption[reddist1]{\label{f:reddist1} These show the contribution of
different redshifts to the mean $y$-distortion over the map, for
the $\Lambda $CDM simulation. In each case the thick line is the
average over 30 separate map realizations shown by the dotted
lines. The upper panel shows the differential contribution per unit
redshift, and the lower one the integrated contribution out to redshift
$z$.}
\end{figure}

In Figure~\ref{f:reddist1} we plot the redshift distribution of the
contributions to the mean $y$-distortion for 30 map realizations in
the $\Lambda$CDM cosmology. This shows the redshifts from which the
bulk of the mean signal originates. Notice the significant scatter
between the individual realizations, because the appearance of bright
clusters at a given redshift occurs primarily due to chance given
their rarity. However the mean contribution is well determined by
averaging over the maps. (We remind the reader that although the maps
are separate realizations, there is only one simulation of each
cosmology so averaging does not completely eliminate cosmic variance.)
We see that the mean SZ signal in the $\Lambda$CDM cosmology comes
from a broad range of redshifts out to around two, and falls off
significantly only beyond that. The signal from nearby is primarily
due to rare but very bright sources, while at large distances it is
due to large numbers of fainter ones.

\begin{figure}
\centering \leavevmode\epsfysize=6.2cm \epsfbox{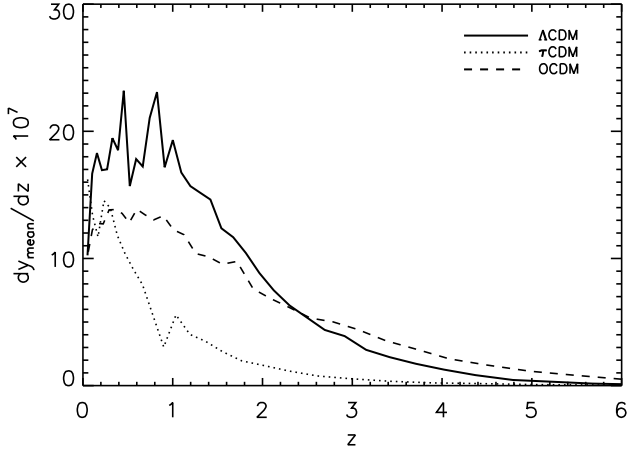}\\
\caption[reddist2]{\label{f:reddist2} The differential contribution
per unit redshift in the three cosmologies, averaged over 30 map
realizations for each.}
\end{figure}

Figure~\ref{f:reddist2} shows the differential redshift distribution,
averaged over the 30 maps, for each of the three cosmologies. The
total mean distortion quoted earlier is the area under the
curves. These results confirm expectation. Close to redshift zero all
the cosmologies give a similar signal, as they must given that they
were normalized to reproduce the present-day cluster temperature
function. In the critical-density case, structure forms latest, and
most of the signal comes from redshifts less than one, while in the
low-density cases the tail extends to much higher redshift. This is
most pronounced in the open case, because structure grows the slowest
there. Finally, we observe that at redshifts around unity, it is the
$\Lambda$CDM cosmology which gives the greatest signal, even though
structures form more slowly in the open case. The reason for this is
that the $\Lambda$CDM cosmology has the greatest volume at these
redshifts and so a larger amount of gas contributes; at redshift one
the physical volume per unit solid angle per redshift interval is 52
percent larger in the $\Lambda$CDM case than for OCDM.

\subsection{The distribution function and source counts}

\begin{figure}
\centering \leavevmode\epsfysize=6cm \epsfbox{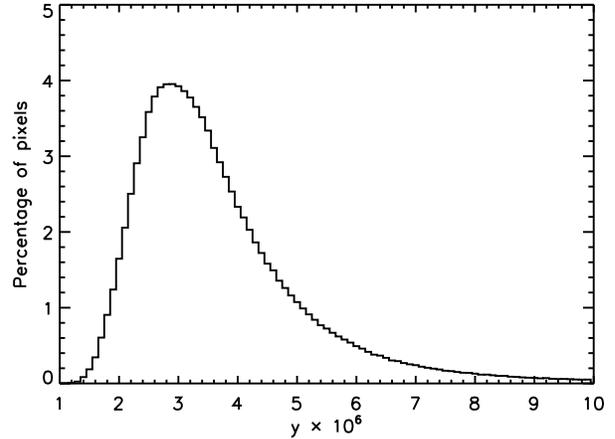}\\ 
\caption[histogram]{\label{f:histogram} A histogram of the $y$ values
in the $\Lambda$CDM maps smoothed to 1\arcmin.}
\end{figure}

In Figure~\ref{f:histogram}, we plot a histogram of the pixel values
in the maps for the $\Lambda$CDM cosmology with 1\arcmin\ smoothing.
This gives the distribution of $y$ values which would be seen were an
instrument of this resolution aimed randomly at the sky. The most
common pixel values are close to the mean y value. In this map the
highest pixel value is just above $2 \times 10^{-5}$.

Of most interest to observers are the source counts above a given flux
level, which determines the detection rate.  We derived these counts
using {\tt SExtractor} \cite{ba96}, a source extraction code based on
a connected-pixel algorithm, which optimally detects, deblends and
measures sources in a given map.  The analysis begins with the iterative 
estimation of the `sky' background were the sources not present. A crucial 
parameter is the threshold level above which sources will be identified. This is 
set as a multiple of the {\em rms} of the background. In our maps, we verified 
that {\tt SExtractor} was able to deblend sources down to a threshold of 
2-sigma; the algorithm therefore becomes source confused at a level roughly 
given by the mean background level $y_{{\rm mean}}$. 

{\tt SExtractor} identifies the location and the total integrated flux
of the sources, though these regions of integration may have
significant overlap and have greatly differing angular extents. We
prefer not to quote the integrated fluxes, which are only appropriate
if the pixels in the map are fully resolved by the observing
apparatus. Instead, to allow for the beam response of different types
of instrument, we detect sources in maps smoothed by gaussians of
different widths, and quote the number of sources with central value $y_{{\rm 
max}}$ above a given
$y$. Recall that the purpose of smoothing the maps is to replace the
precise value of $y$ at a given point with the value of $y$ which
would be seen by an observational beam aimed at that point. The number
of sources which have $y_{{\rm max}}$ exceeding the instrument
threshold is therefore precisely the number of sources that could be
detected by a complete scanning of that field by the instrument. Our
smoothings correspond to idealized perfectly-gaussian beams, and we
have verified visually that {\tt SExtractor} performs well on our
maps.

\begin{figure}
\centering \leavevmode\epsfysize=5.8cm \epsfbox{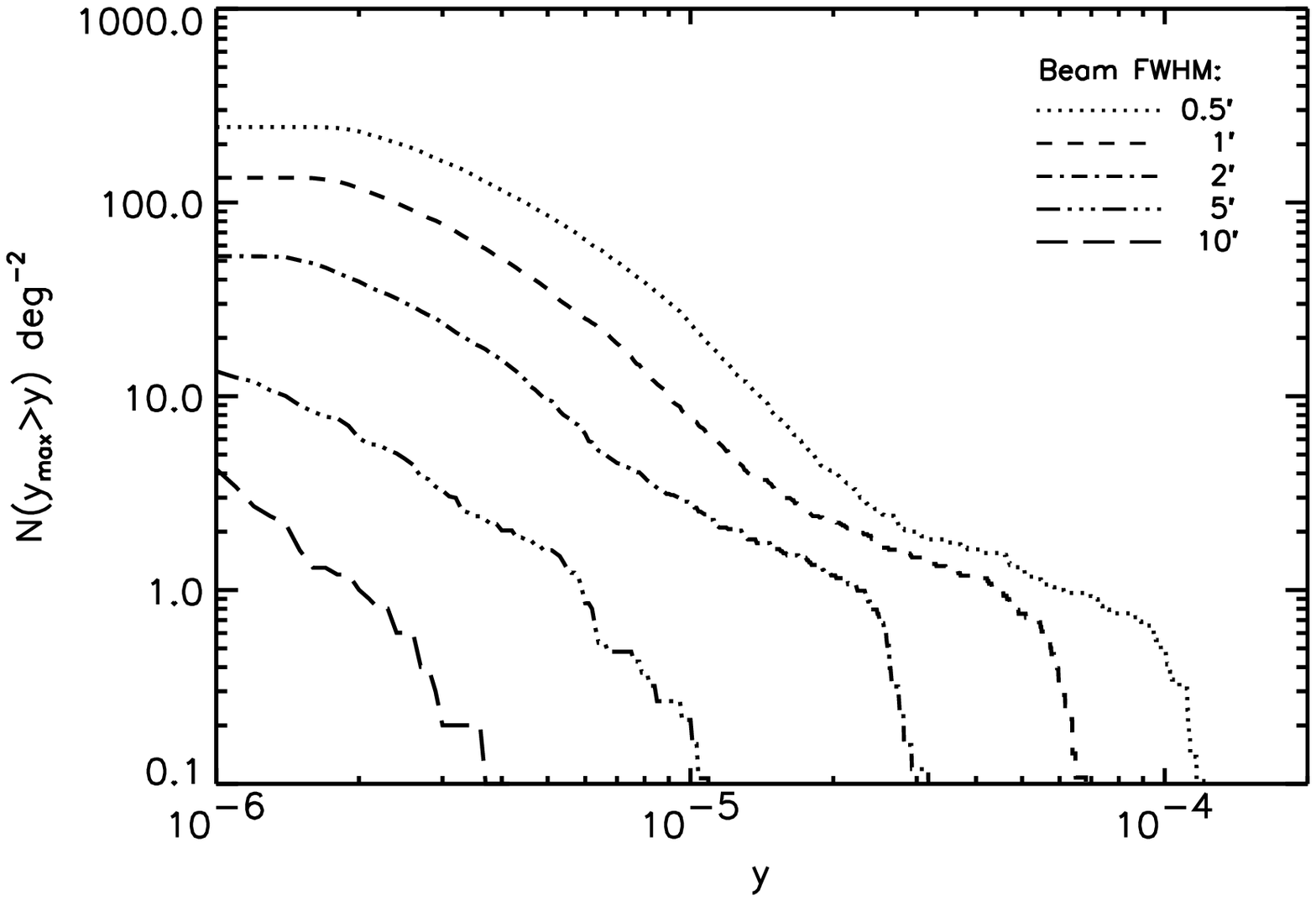}\\ %FLAT
\centering \leavevmode\epsfysize=5.8cm \epsfbox{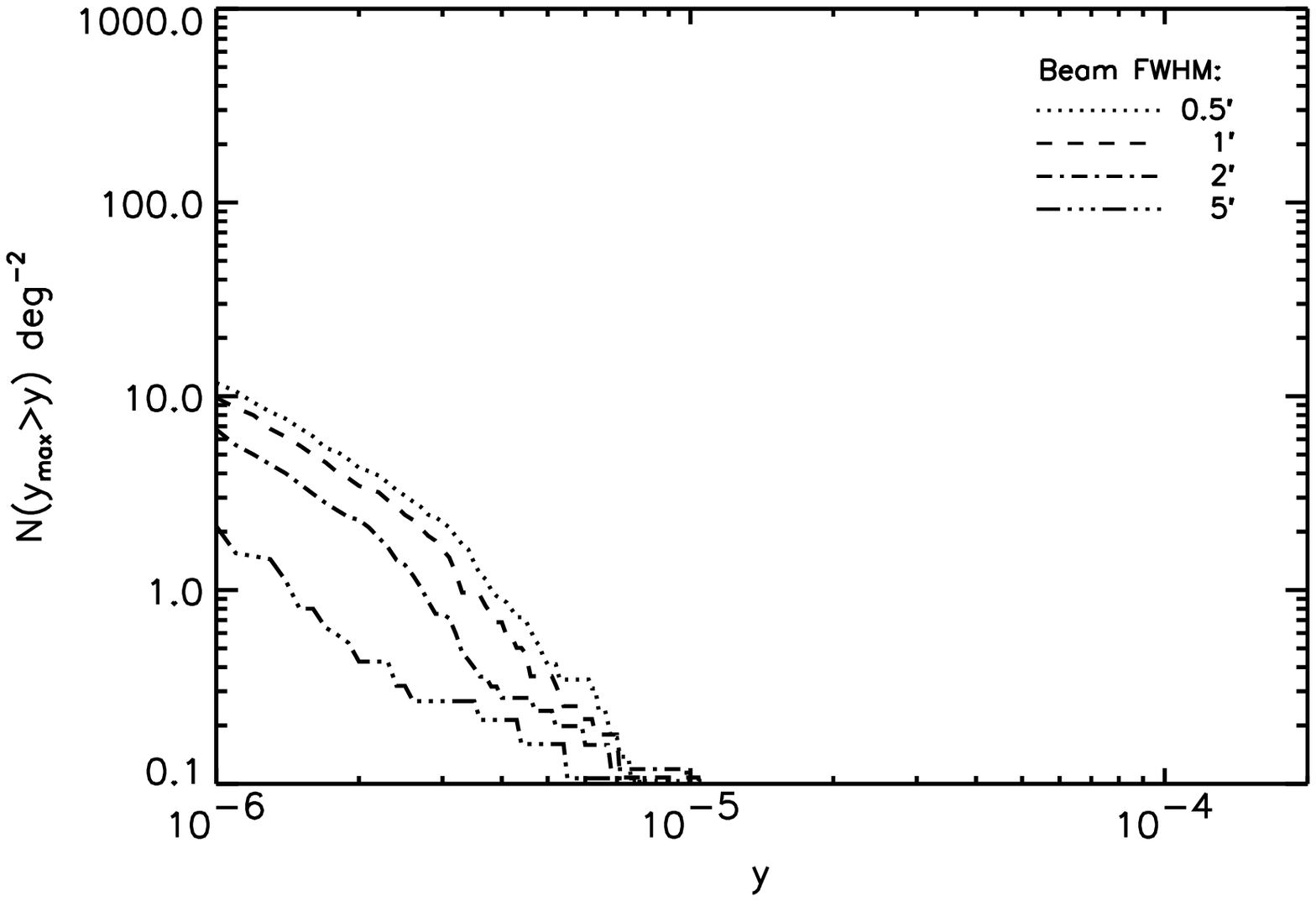}\\ %CRIT
\centering \leavevmode\epsfysize=5.8cm \epsfbox{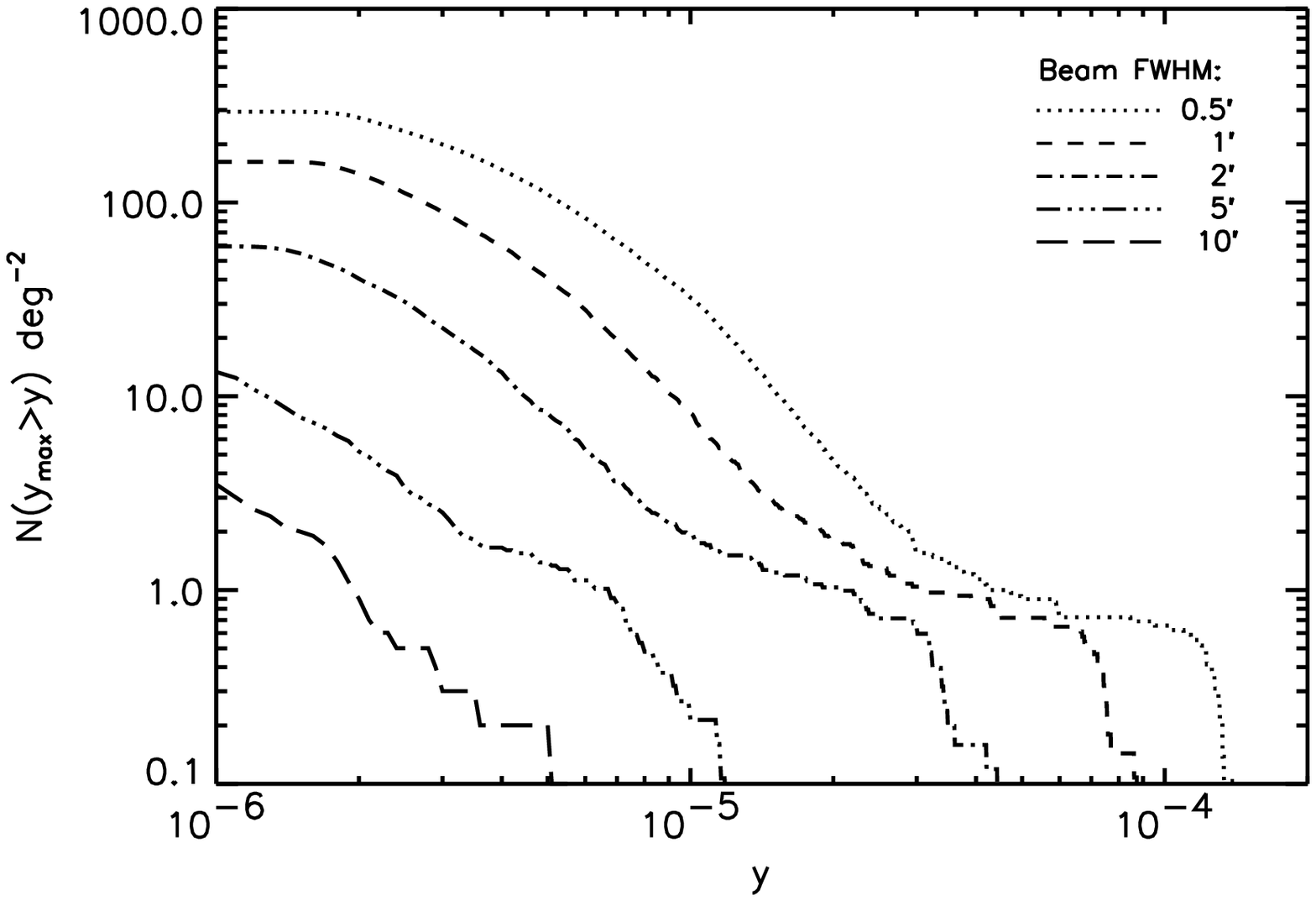}\\ %OPEN
\caption[counts]{\label{f:counts} Source counts for each cosmology,
averaged over 30 map realizations. From top to bottom they are
$\Lambda$CDM, $\tau$CDM and OCDM. The lines correspond to map smoothings by 
gaussians 
with FWHM of 0.5\arcmin, 1\arcmin, 2\arcmin, 5\arcmin\ and 10\arcmin.  
}
\end{figure}

Figure~\ref{f:counts} shows the source counts for the three different
cosmologies, per square degree on the sky, with maximum distortion
exceeding a given $y$. A range of different smoothings are shown,
corresponding loosely to different types of instrument; for example,
BIMA and SuZie both have beams with FWHM around 1.5\arcmin, while {\sc
Planck} ranges from 5\arcmin\ to 10\arcmin\ for the channels most
sensitive to the thermal SZ effect. At the highest values of $y$ the
predictions become uncertain as the low number of sources means there
is significant cosmic variance; although we average over many map
realizations we have only a single hydrodynamical simulation for each
cosmology. The curves flatten out at low values of $y$ as the 
{\tt SExtractor} threshold is approached and the routine becomes source 
confused; we see that this happens around $y_{{\rm mean}}$, with some 
improvement when the
maps are highly smoothed as the {\tt SExtractor} threshold decreases with 
smoothing. Between these two limiting regions, the
source counts are well described by power-laws, with $d \ln N(y_{{\rm
max}}>y)/d\ln y \simeq -2$ in all cases.

\begin{figure}
\centering \leavevmode\epsfysize=5.8cm \epsfbox{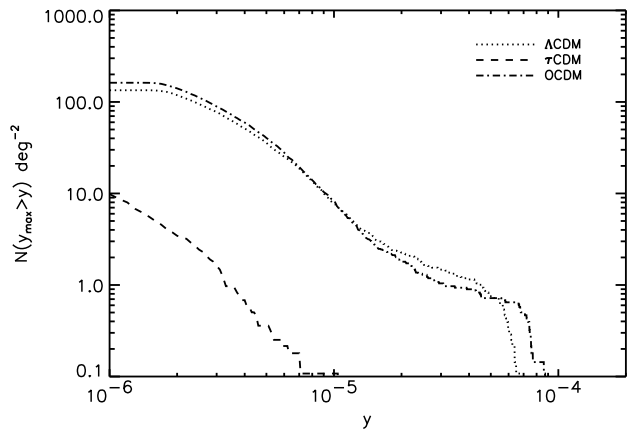}\\ 
\caption[counts2]{\label{f:counts2} Source
counts, with 1\arcmin\ smoothing, for each cosmology. At $y =
10^{-5}$, there are about 80 times fewer sources in the
critical-density case as compared to the two low-density cases.}
\end{figure}

In Figure~\ref{f:counts2}, we show all three cosmologies together, for
the 1\arcmin\ smoothing. We see that across the entire reliable range
the critical-density case falls well below the two low-density ones,
which are very close to one another. The difference is more than a
factor of ten even at the bright end. This is because although the
simulations are normalized to give comparable contributions at
redshift zero, in low-density models the brightest SZ clusters are
visible to fairly high redshifts, due to the redshift-independent
surface brightness of resolved SZ sources.

The bump at the bright end for the low-density models, coming from the
largest cluster in the simulation box, is an indication of the
limitations of having only a single simulation; it does not disappear when
we average over maps made from that realization. Note that this
feature is shared by the two low-density models because they are run
from the same initial conditions, not because it is a genuine feature.

Of obvious interest is the comparison between the simulation results and 
the theoretical predictions based on the Press--Schechter prescription 
(Press \& Schechter 1974). Detailed calculations (Barbosa et al., in 
preparation) 
show good agreement between the results obtained by both methods for the 
source counts distribution and the global $y$ distortion. Earlier 
Press--Schechter calculations (Barbosa et al.~1996) predicted a substantial 
mean $y$, close to the FIRAS limit for a low-density cosmological model. The 
difference lies solely in the way the power spectrum was normalized to the 
present abundance of galaxy clusters. In the present work, the shape of the 
power spectrum was chosen to give the best fit to the observed
galaxy power spectrum ($\Gamma=0.21$). In Barbosa et al.~(1996), 
both the normalization and the shape of the power spectrum were allowed to be 
determined solely by the cluster abundances. Although the normalizations 
were similar, the differences in the power spectrum shape enhanced 
the abundances of smaller structures such as groups and small clusters, which 
appear to contribute the most to the mean $y$ distortion (Hernandez-Monteagudo, 
Atrio-Barandela \& M\"ucket 2000).

In studies for the {\sc Planck} HFI instrument (Puget et al.~1998; Hobson et
al.~1998), it has been estimated that maximum entropy methods would allow a
cluster with a central $y$ of $3 \times 10^{-6}$ to be picked out against other
contributions (i.e.~dust emission, point sources and the cosmic microwave
background fluctuations), at a FWHM resolution of 5\arcmin.  However one cannot
use this literally, without taking into account that we find a mean $y$ which is
comparable.  With {\sc Planck}, each detector measures the fluctuations with
respect to the mean seen by that detector, so that {\it at each frequency} the
mean intensity is unobservable.  {\sc Planck} therefore cannot see the mean 
$y$-distortion, and
the quoted sensitivity for cluster detection must be interpreted as the level
above the mean.\footnote{Hence in principle {\sc Planck} might measure negative
$y$ values in some directions.}  This has the unfortunate effect of reducing the
difference in the number of sources between the low-density and critical-density
cases, because having more clusters raises the mean signal too.  We estimate 
that for
the favoured $\Lambda$CDM cosmology, there are order 0.6 sources per
square degree of greater than this brightness, implying a total number across
the sky of order 25000.  For the critical-density model, the estimate
is a factor of 3 less.  These estimates are in reasonable
agreement with those made so far (Haehnelt 1997; Aghanim et al.~1997; Puget et
al.~1998).

\section{Conclusions}

Studies of the Sunyaev--Zel'dovich effect are reaching observational
maturity, and detailed simulations are required to interpret upcoming
data. We have used hydrodynamical $N$-body simulations to construct
maps of the thermal SZ effect for three different cosmological
models. This has enabled us to study a range of properties, including
the mean $y$-distortion averaged over the maps and the source counts
expected in the different cosmologies at a series of different angular
resolutions, including that of the {\sc Planck} satellite.

Although clusters are the main contributors to the visible SZ signal, it has
been suggested (e.g.  see Refregier, Spergel \& Herbig 2000 and references
therein) that the filamentary structures containing the 
majority of baryons
could also be important.  Our simulations seem to show that detecting the 
filamentary SZ
effect is challenging for upcoming experiments --- there are no
obvious filaments in Figure~\ref{f:maps}.  However, first of all note that our
maps correspond to randomly-chosen areas of the sky, whereas a search for the
filamentary SZ effect would naturally be focussed initially on nearby known
large filaments.  More importantly, as filaments constitute the birthing pools
of galaxies, non-gravitational heating (not included in our simulations)
injected into the intergalactic medium could raise the temperature of the 
filamentary gas sufficiently to produce
significant spectral distortions in the CMB (Cen \& Ostriker 1992; Refregier et
al.~2000; Valageas \& Silk 1999).  (This will also increase the mean 
$y$-distortion, potentially moving
it close to the present FIRAS limit.)  This may be beneficial, because CDM, as a
hierarchical structure formation family of models, can lead to the
overproduction of small structures such as galaxy halos and possibly groups,
structures corresponding to the lower $y$ values visible in the simulations.
Inclusion of non-gravitational heating can suppress baryon infall into the
smallest dark matter potentials (Cole 1991; Blanchard, Valls-Gabaud \& Mamon
1992; Navarro \& Steinmetz 1997).  Heating may therefore lead to a decrease
in the number of sources producing small $y$ values.

Our principal results are the following. The mean $y$-distortion is
around $4 \times 10^{-6}$ for low-density cosmologies and $1 \times
10^{-6}$ for critical density. In the low-density cosmologies, it is
contributed across a broad range of redshifts, with the bulk coming
from $z \la 2$ and a tail out to $z \sim 5$, while for
critical-density models most of the contribution comes from $z < 1$.
The number of SZ sources above a given $y$ depends strongly on
instrument resolution. For a 1\arcmin\ beam, there is around one
source per square degree with $y > 10^{-5}$ in a critical-density
Universe, and around 10 such sources per square degree in low-density
models. Low-density models with and without a cosmological constant
give very similar results. We estimate that the {\sc Planck} satellite
will be able to see of order 25000 SZ sources if the Universe has a
low density, or around 10000 if it has critical density.

%%%%%%%%%%%%%%%%%%%%%%%%%%%%%%%%%%%%%%%%%%%%%%%%%%%%%%%%%%%%%%%%%%%%%%
\section*{Acknowledgments}

We are indebted to Hugh Couchman and Frazer Pearce for their part in
writing the hydrodynamical $N$-body code used to generate the
simulation data used in this work. We thank Anthony Lasenby for important 
discussions regarding the {\sc Planck} satellite's inability to measure the mean 
spectrum, and John Carlstrom, Ian
Grivell, David Spergel and Aprajita Verma for helpful discussions and
comments. ACdS was supported by FCT (Portugal), DB by the European
Union TMR programme, ARL in part by the Royal Society, and PAT by a PPARC 
lecturer fellowship. 
We
acknowledge use of the Starlink computer systems at Imperial College
and at Sussex. The simulations were carried out on the BFG--HPC
facility at Sussex funded by HEFCE and SGI, and part of the data
analysis on the {\sc cosmos} National Cosmology Supercomputer funded
by PPARC, HEFCE and SGI.
%%%%%%%%%%%%%%%%%%%%%%%%%%%%%%%%%%%%%%%%%%%%%%%%%%%%%%%%%%%%%%%%%%%%%%

%%%%%%%%%%%%%%%%%%%%%%%%%%%%%%%%%%%%%%%%%%%%%%%%%%%%%%%%%%%%%%%%%%%%%%

\bsp

\end{document}